\documentclass[twocolumn,twoside,slac_two,floatfix]{revtex4}
\usepackage{graphicx}
\usepackage{fancyhdr}
\newcommand{\ve}[1]{\mbox{\boldmath$ #1 $}}

\begin{document}

\title{Radiative Effect on Particle Acceleration via Relativistic Electromagnetic Expansion}

%

\author{K. Noguchi\footnote{noguchi2@llnl.gov}, E. Liang}
\affiliation{Rice Univ. TX 77005 USA}
%

\begin{abstract}
The radiation damping effect on the diamagnetic relativistic pulse accelerator
(DRPA) is studied in two-and-half dimensional Particle-in-Cell (PIC) simulation
with magnetized electron-positron plasmas.
Self-consistently solved radiation damping force converts particle energy
to radiation energy. 
The DRPA is still robust with radiation, and the Lorentz factor of the most
high energy particles reach more than two thousand before they decouple from
the electromagnetic pulse.
Resulted emitted power from the pulse front is lower in
the radiative case than the estimation from the non-radiative 
case due to the radiation damping.
The emitted radiation is strongly linearly polarized and 
peaked within few degrees from the direction of Poynting flux.

\end{abstract}

\maketitle

\thispagestyle{fancy}


\section{INTRODUCTION\label{sec:sec1}}
Radiation loss and damping can become important in the plasma energetics and 
dynamics when charged particles suffer extreme acceleration. In the 
ultra-relativistic regime, the accumulated effect from radiation damping can 
severely limit individual particle acceleration even if the radiation damping 
force is weak compared to external forces. However, conventional Particle-in-Cell 
simulations of collisionless plasmas have not included radiation effects. 
We developed a new 2-1/2-D code including self-consistent radiation damping, 
and studied a effect of radiation damping in particle acceleration driven by 
relativistic pulse accelerator (DRPA)~\cite{lian03} in the strong magnetic field 
limit. From the radiation-damped plasma and field evolution, we obtained the
observable high energy radiation output. 
This radiative PIC simulation code is applicable  
to to a wide range of high-energy astrophysics 
phenomena (pulsars, blazars, gamma-ray bursts) and ultra-intense laser 
applications ~\cite{zhid02}. 

The outline of this paper is as follows. In Sec.~\ref{sec:sec2}, the derivation
of the radiation damping force in the relativistic form is shown. In Sec.~\ref{sec:sec3}
we explain how the radiation damping force is implemented into the PIC code.
Results of simulation and comparison with non-radiative case are given in 
Sec.~\ref{sec:sec4}, and we summarize in Sec.~\ref{sec:sec5}. 
 
\section{DERIVATION OF RADIATIVE FORCE\label{sec:sec2}}
In the conventional PIC simulations, radiation damping force $\ve{f}_{\!rad}$ is ignored 
because of its small amplitude compared to external forces $\ve{F}_{\!ext}$. 
However, it is not negligible when
the accumulated work done by the radiation damping during deceleration time $\tau_{decl}$
is comparable to the work done by the external force in typical acceleration time 
~\cite{zhid02}.

It is impractical to include high-frequency radiation wave into the electromagnetic 
calculation in the PIC code, since the typical radiation wavelength is much shorter 
than the spatial resolution of the fields ($\sim$ Debye length
$\lambda_D \equiv c/\omega_{pe}$, where $\omega_{pe}=\sqrt{4\pi\rho e/m_e}$ is the 
electron plasma frequency).
Accelerated particles can emit up to the critical frequency $\omega_c=3\gamma^2\Omega_{ce}$,
where $\gamma=E/m_e c^2=1/\sqrt{1-v^2/c^2}$ is the Lorentz factor 
and $\Omega_{ce}=eB/(m_e c)$ is the electron gyro-frequency.
The ratio of the critical radiation wavelength $\lambda_c$ to $\lambda_D$ is given by
$\lambda_c/\lambda_D=(2\pi\omega_{pe})/(3\gamma^3\Omega_{ce})$, which is $\ll 1$
because $\omega_{pe}/\Omega_{ce}<0.1$ in magnetic-dominated cases and $\gamma\gg 1$.

Instead, we introduce a radiation damping force in the form of the Dirac-Lorentz 
equation~\cite{land75}, which is proven to be exact for a classical 
point particle~\cite{rohr01}. The 4-vector form of Dirac-Lorentz equation with 
the damping force $g^i$ is given by
\begin{equation}
mc\frac{du^i}{ds}=\frac{e}{c}F^{ik}u_k+g^i,\label{four}
\end{equation}
where $F^{ik}$ is the electromagnetic tensor~\cite{jack75}, $u^i$ is the velocity 
four-vector, and $g^i$ is the radiation damping force
\begin{equation}
g^i=\frac{2e^2}{3c}\left(\frac{d^2u^i}{ds^2}-u^iu^k\frac{d^2u_k}{ds^2}\right),\label{radf}
\end{equation}
which satisfies the auxiliary relation $g^iu_i=0$. 
Since Eq. (\ref{radf}) includes the second derivative of $u^i$, there exist 
unphysical run-away solutions~\cite{jack75}. 
To avoid the run-away solutions, we assume $g_i\ll eF^{ik}u_k/c$ and eliminate 
the second derivative terms in $g^i$ by expressing 
$d^2u^i/ds^2$ with the first derivative of Eq. (\ref{four}) with $g^i=0$,
\begin{equation}
\frac{d^2u^i}{ds^2}=\frac{e}{mc^2}\frac{\partial F^{ik}}{\partial x^l}u_ku^l
+\frac{e^2}{m^2c^4}F^{ik}F_{kl}u^l.
\end{equation}
Using the fact $(\partial F^{ik}/\partial x^l)u_iu_k=0$, we find
\begin{eqnarray}
g^i=\frac{2e^3}{3mc^3}\left[\frac{\partial F^{ik}}{\partial x^l}u_ku^l
-\frac{e}{mc^2}F^{il}F_{kl}u^k\right.\nonumber\\
\left.+\frac{e}{mc^2}(F_{kl}u^l)(F^{km}u_m)u^i\right].
\end{eqnarray} 
The resulting damping force term $\ve{f}_{rad}$ in the 3-vector form is 
given by~\cite{land75}
\begin{eqnarray}
\ve{f}_{rad}\!\!&=&\frac{2e}{3\Omega_{ce}}k_{rad}\times\nonumber\\
&&\!\!\!\!\!\!\!\!\left\{\gamma\left[\left(\frac{\partial}{\partial t}+\ve{v}\cdot\nabla\right)\!\!\ve{E}
+\frac{\ve{v}}{c}\times\!\!\left(\frac{\partial}{\partial t}+\ve{v}\cdot\nabla\!\!\right)\!\!\ve{B}\right]\right.\nonumber\\
&&\!\!\!\!\!\!\!\!+\frac{e}{mc}\left[\ve{E}\!\times\!\ve{B}+\frac{1}{c}\ve{B}\!\times\!(\ve{B}\!\times\!\ve{v})
+\frac{1}{c}\ve{E}(\ve{v}\cdot\ve{E})\right]\nonumber\\
&&\!\!\!\!\!\!\!\!\left.-\frac{e\gamma^2}{mc^2}\ve{v}
\left[\left(\ve{E}+\frac{1}{c}\ve{v}\times\ve{B}\right)^{\!\!2}\!\!-\frac{1}{c^2}(\ve{E}\cdot\ve{v})^2\right]\right\},\hspace{3mm}
\label{radf2}
\end{eqnarray}
where $\ve{v}$ is the 3-velocity, and $\ve{E}$ and $\ve{B}$ are the self-consistent 
electric and magnetic fields. Here we introduce a non-dimensional 
factor $k_{rad}$ given by
\begin{equation}
k_{rad}=\frac{r_e\Omega_{ce}}{c}=1.64\times10^{-16}\times B\mbox{(gauss)},
\end{equation}
where $r_e=e^2/(mc^2)$ is the classical electron radius. 

The first term of the radiation damping force (\ref{radf2}) represents the radiation 
damping due to the ponderomotive force acceleration. 
The third term is Compton scattering by large scale $(\lambda >\lambda_D)$ electromagnetic field
which reduces to Thomson scattering in the classical limit~\cite{rybi79}. 
Note here that the scattering between radiation field and particles is not 
considered since electromagnetic field in Eq. (\ref{radf2}) is averaged over 
the grid separation that is much larger than the wavelength of radiation, as we discussed above.
In other words, plasma is perfectly transparent to emitted radiation, and
all the radiation disappears from the simulation box immediately after particles radiate.

The magnetic field strength of pulsars are in the 
range of $10^{-4}\leq k_{rad}\leq 10^{-3}$, and $k_{rad}\simeq 10^{-2}$ for magnetars. 
In the laboratory, $k_{rad}\simeq
10^{-6}$ $(I/10^{22}$W{}cm$^{-2})^{1/2}$ for lasers of intensity $I$. 
Hence for ultraintense laser interactions
($I\geq 10^{22}$W{}cm$^{-2}$), radiation damping effects can become significant 
on simulation time scales $>0.3$ps ($=10^5\Omega_{ce}^{-1}$). (See also~\cite{zhid02})
In our simulation, we can enhance the radiation effect by increasing 
the value of $k_{rad}$, or the 'effective' electron radius. 
However, we should restrict ourselves not to reach the quantum-limit, 
$\hbar\Omega_{ce}\sim m_e c^2$ or $B>4.4\times 10^{13}$ gauss, which corresponds 
to $k_{rad}=7.2\times10^{-3}$, or the Dirac-Lorentz equation (\ref{four}) fails.
We choose $k_{rad}$ from zero to $10^{-3}$ in the simulation to enhance the radiation 
effect and $|\ve{f}_{rad}|\tau_{sim}\simeq|\ve{F}_{ext}|\Omega_{ce}^{-1}$ so we can 
see the difference between radiative and non-radiative ($k_{rad}=0$) case within 
the simulation time-scale $\tau_{sim}\simeq10^4\Omega_{ce}^{-1}$.

\section{IMPLIMENTATION OF RADIATION FORCE\label{sec:sec3}}
The 2-1/2D explicit PIC simulation scheme is used 
with the explicit leap-frogging method for time advancing~\cite{bird85}. 
Spacial grids for the fields are uniform in both $x$ and $z$ directions, 
$\Delta x=\Delta z=\lambda_D$. The simulation domain in the $x\!-\!z$ plane is
$-L_{x}/2\leq x\leq L_{x}/2$ and
$0\leq z\leq L_{z}$ with a doubly periodic boundary condition in both directions.

Following Liang et.~al.~\cite{lian03}, the initial plasma is uniformly distributed 
at the center of the simulation box, $-6\Delta x<x<6\Delta x$ and $0<z<L_z$. 
The background uniform magnetic 
field $\ve{B}_0=(0,B_0,0)$ is applied only in the same region, so that the magnetic field freely 
expands toward the vacuum regions, $x>6\Delta x$ and $x<-6\Delta x$ with accelerating plasma. 
We choose $L_x$ to be long enough so that plasma and EM wave never hit the boundaries 
in the $x$ direction within the simulation time.

The initial temperature of plasma is assumed to be a spatially uniform relativistic Maxwellian, 
$k_BT_e=k_BT_p=1$MeV, where the subscripts $e$ and $p$ refer to electrons and positrons. 

The radiation damping force (\ref{radf2}) is calculated self-consistently and 
fully-explicitly as follows. The velocity of each particle should be updated each time step
 from $\ve{v}(t-\Delta t/2)$ to $\ve{v}(t+\Delta t/2)$, using the electromagnetic field
and $\ve{f}_{rad}$ at time $t$ following the equation of motion in the relativistic form,
\begin{equation}
\frac{d\ve{p}}{dt}=e\ve{E}+\frac{e}{c}\ve{v}\times\ve{B}+\ve{f}_{rad},
\end{equation}
where $\ve{p}=m\gamma\ve{v}$ is the relativistic momentum. 
All the terms in the equation (\ref{radf2}) are given by
the ordinary leap-frog field solver except $d\ve{E}(t)/dt$ term, 
which is not calculated at time $t$ in the conventional PIC code. 

In order to calculate $d\ve{E}(t)/dt$, we need to know $\ve{J}(t)$. 
First, we advance the position $\ve{x}$ half timestep 
from $t-\Delta t/2$ to $t$ for each particle using 
velocity $\ve{v}(t-\Delta t/2)$. Next, temporal current $\ve{J}_t(t)$ is calculated from 
$\ve{v}(t-\Delta t/2)$ and $\ve{x}(t)$. Note that this current $\ve{J}_t(t)$ is not exact
because of the velocity field not at time $t$ but $\ve{v}(t-\Delta t/2)$.
Finally, the term $d\ve{E}(t)/dt$ is calculated using the Maxwell equation, and the 
radiation damping 
force is calculated for each particle.
To update the velocity, we consider a sum of the radiation force and the electric field force 
as a net 'acceleration' force, and apply the Boris method for particle gyration~\cite{bird85}.

\section{RESULTS\label{sec:sec4}}

\begin{table}
  \caption{\label{table1}Parameters for each runs}
    \begin{tabular}{c|c|c|c}
      &$k_{rad}$&$\omega_{pe}/\Omega_{ce}$& Duration $t\Omega_{ce}$\\
      \hline
      Run A & 0         & 0.1 & 10000\\
      Run B & $10^{-4}$ & 0.1 & 10000\\
      Run C & $10^{-3}$ & 0.1 & 10000\\
      Run D & 0         & 0.01 & 70000\\
      Run E & $10^{-4}$ & 0.01 & 70000\\
      Run F & $10^{-3}$ & 0.01 & 70000\\
    \end{tabular}
\end{table}

We choose six different sets of parameters shown in Table~\ref{table1}, by changing 
$k_{rad}=0$, $10^{-4}$, $10^{-3}$ and $\omega_{pe}/\Omega_{ce}=0.1$, $0.01$, 
and run simulations for each case.
Hereafter, we call Run A-C as the weak magnetic field case and Run D-F as 
the strong magnetic field case, based on the ratio $\omega_{pe}/\Omega_{ce}$.

\begin{figure}[floatfix]
\includegraphics[width=\linewidth]{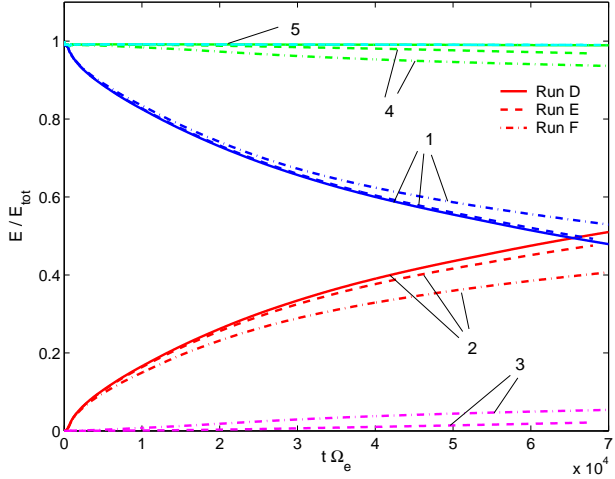}
\caption{\label{fig:eng} System-integrated energy in the electromagnetic field (1), 
particles (2), radiation damping (3), sum of field and particle energy (4), and total
energy (5) functions of time for Run D (solid), E (dashed) and F (dash-dot).}
\end{figure}

First, we check the total energy conservation for Run D,E  and F in Fig.~\ref{fig:eng}. 
In the radiative (RD) cases, 
the energy loss by the radiation $E_{rad}$  is obtained from the time integral
\begin{equation}
E_{rad}(t)=\int_0^t\left(\sum_{e,p}\ve{v}(t')\cdot\ve{f}_{rad}(t')\right) dt'\label{Erad}.
\end{equation}
In the non-radiative (NRD, $k=0$) case, total energy of the system is given by the sum of the field energy $E_{fie}$ (Line 1) and kinetic energy $E_{kin}$ (Line 2), and it conserves. 
In the RD case, however, sum of $E_{kin}$ and $E_{fie}$ does not 
conserve (Line 4), but sum of $E_{kin}$, $E_{fie}$ and the radiation energy $E_{rad}$ (Line 3) conserves (Line 5),
indicating that the radiation damping force is self-consistently calculated. 
In all the RD cases, the energy is transferred from field to particle, and then radiation, 
indicating the DRPA mechanism accelerates particles even in the RD cases.
Energy transfer, however, from field to particles becomes less efficient 
with larger radiation damping force. Radiation prevents energetic particles to
get accelerated from EM field, resulting less efficiency of transfer in high energy tail of
the particle distribution.

\begin{figure}
\includegraphics[width=\linewidth]{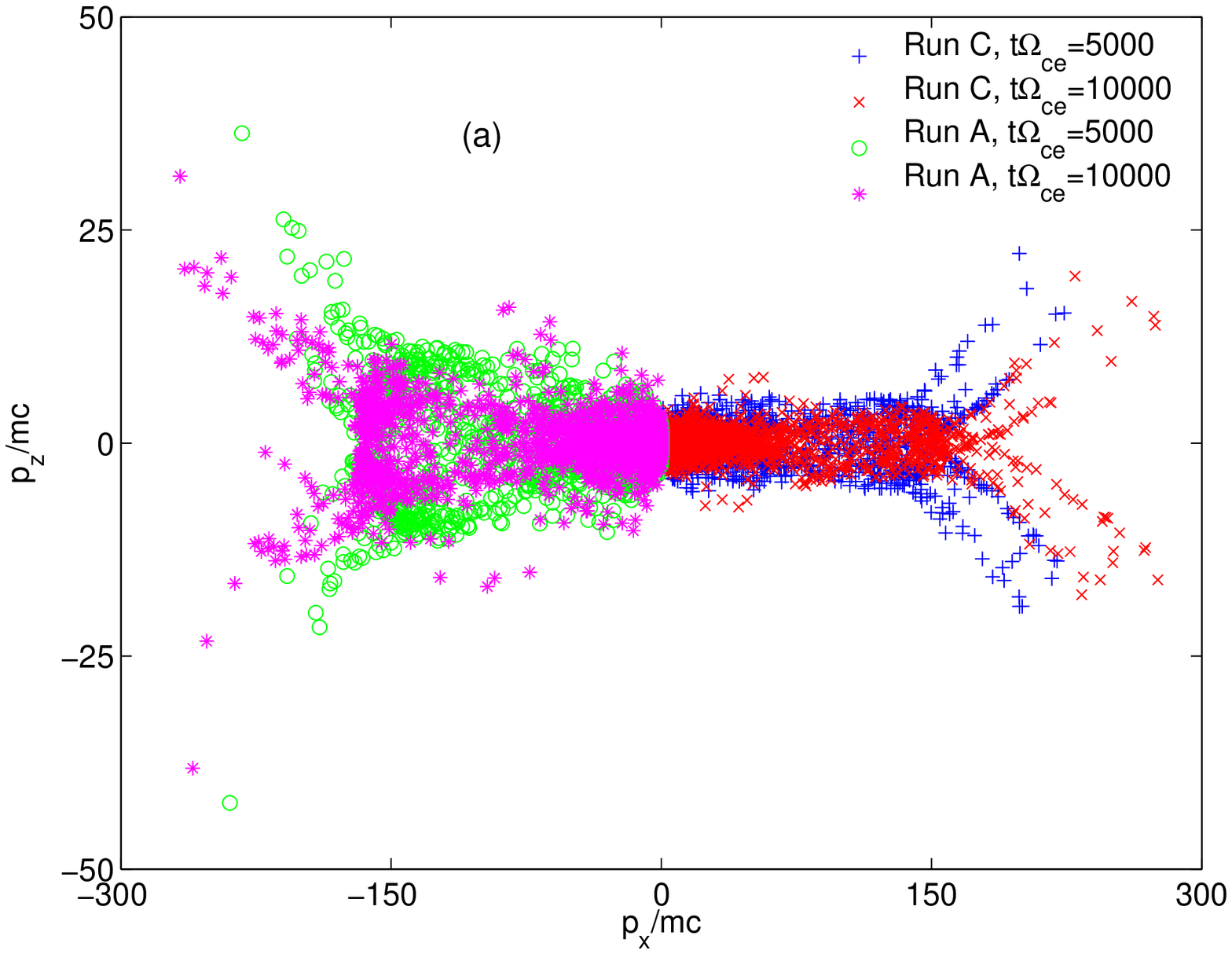}
\includegraphics[width=\linewidth]{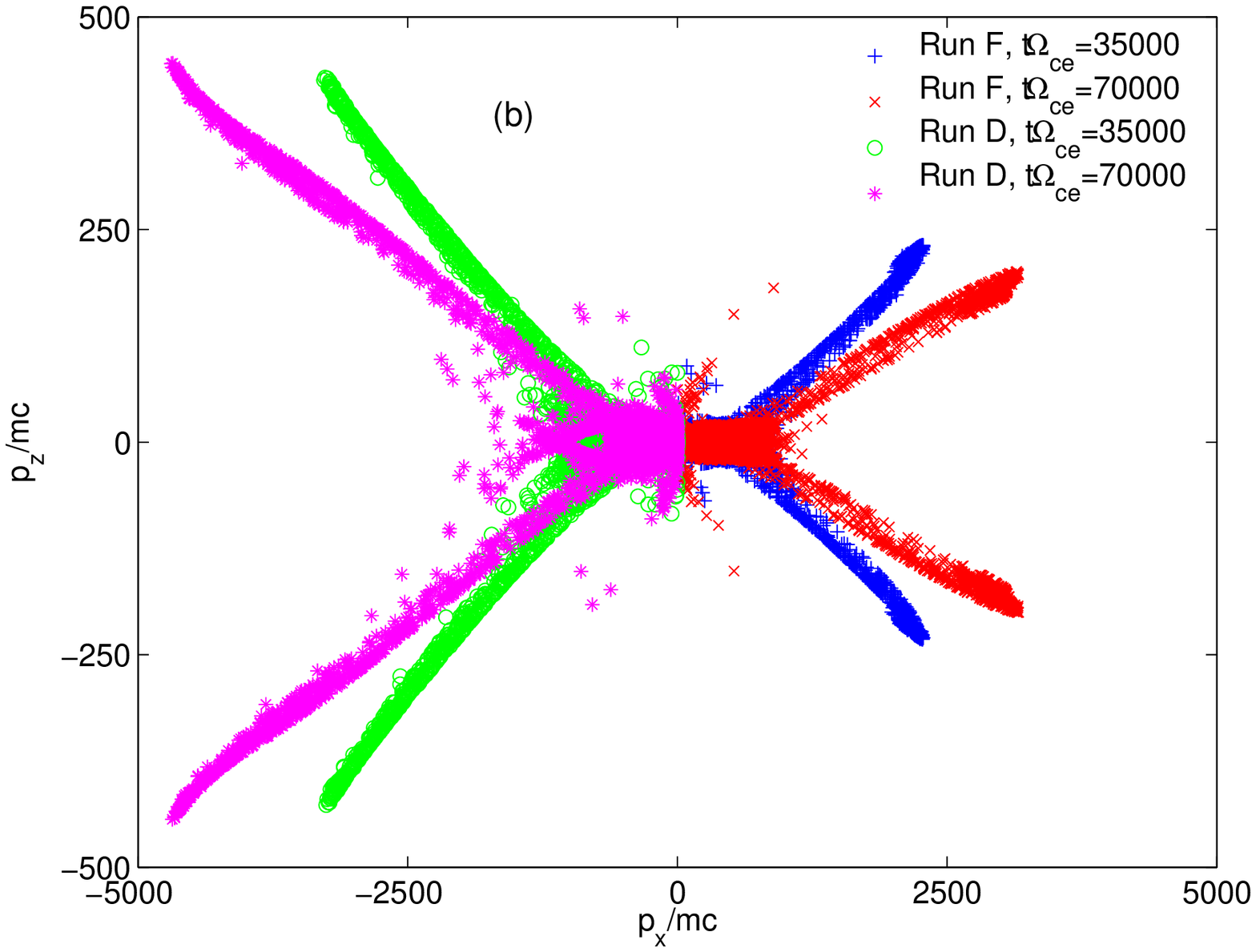}
\caption{\label{fig:phs}Momentum distribution of particles for Run A [(a), $x<0$] and Run C [(a), $x>0$]
at $t\Omega_{ce}=5000$ (green, blue) and $t\Omega_{ce}=10000$ (magenta, red), and
Run .D [(b), $x<0$] and Run F [(b), $x>0$] at $t\Omega_{ce}=35000$ (green, blue) and $t\Omega_{ce}=70000$
(magenta, red). Results in the positive and negative $x$ directions are identical in all cases.}  
\end{figure}

Figure \ref{fig:phs} shows the momentum distribution of particles for (a) Run A($x<0$) and Run C($x>0$), 
and (b) Run D($x<0$) and Run F($x>0$). 
The DRPA accelerates electrons and positrons in the same direction along the $x$ axis, 
whereas electric field $E_z$ accelerates electrons in the positive $z$ and positrons
in the negative $z$ direction respectively in the positive $x$ region. In the negative $x$
region, acceleration directions are opposite for both species, forming X shape distribution
in the $p_x-p_z$ plane as a result. 
Charge separation does not occur because of the periodicity 
in the $z$ direction. Resulted induced current $J_z$ accelerates particles in the $x$ 
direction by the ponderomotive force $\ve{J}\times\ve{B}$. The ponderomotive force creates
successive 'potential wells' in the $x$ direction, which captures and accelerates co-moving 
particles. We emphasize here that there is no charge separation in the $x$ direction
because of no mass difference between electron and positron.

Obviously, particle momenta in both $x$ and $z$ directions are radiated away in the RD cases
in both weak and strong magnetic field cases. 
Bifurcation in high energy tails occurs in Run C since slow particles
can not keep up with the speed of the first (and the fastest) ponderomotive potential well and
decoupled from it. Then the second potential well captures these slow particles and accelerates
them again, which creates the bifurcated high energy tails in the phase space.
The bifurcation is less clear in Run A than Run C because energetic particles 
are not suffered from the radiation damping and most of them can stay in the 
first potential well. 

In the strong magnetic field case [Fig.~\ref{fig:phs}(b)], acceleration is strongly reduced
in the RD case (Run F). For high energy ($\gamma\gg1$) particles, the Compton scattering 
[the third term in Eq. (\ref{radf2})]
becomes a dominant term because of the large Lorentz factor, 
and makes the DRPA acceleration less efficient. Bifurcation is strongly 
suppressed in both Run D and F, since the first potential well is deep enough to
capture almost all of the energetic particles.

\begin{figure}
\includegraphics[width=\linewidth]{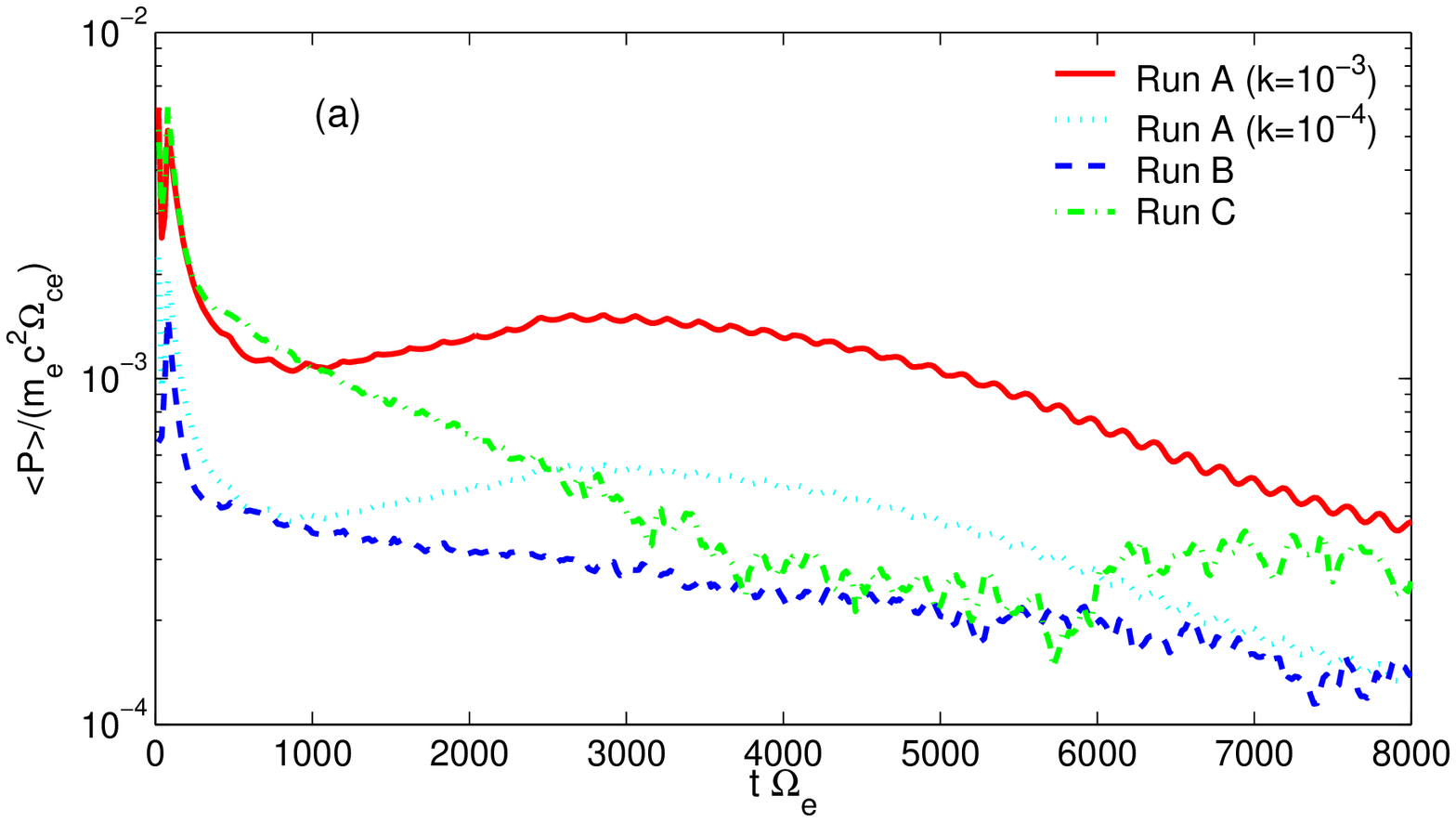}
\includegraphics[width=\linewidth]{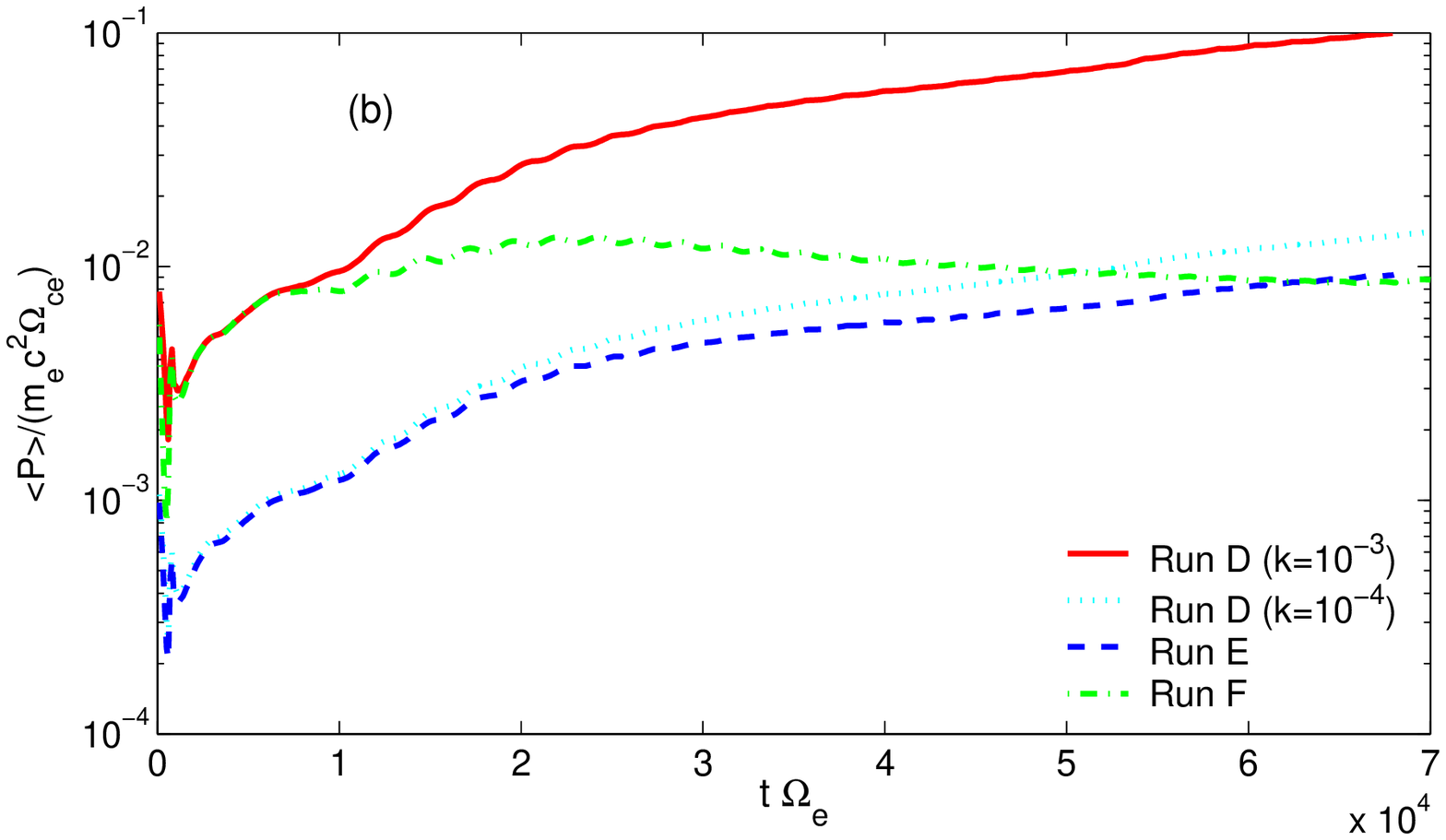}
\caption{\label{fig:pow} Average instantaneous radiation power from particles within $30\lambda_D$ 
from pulse front for (a) $\omega_{pe}/\Omega_{ce}=0.1$ and (b) $0.01$ with $k_{rad}=1-^{-3}$ (dash-dot line) and $k_{rad}=10^{-4}$ (dashed line). 
Estimated power for NRD cases with $k_{rad}=10^{-3}$ (solid line) and $10^{-4}$ (dotted line) are also shown for comparison.}
\end{figure}

Next, we compare the radiation power of RD cases with NRD cases.
For the NRD case, we estimate the radiation power using the relativistic dipole formula~\cite{rybi79}
\begin{equation}
\langle P\rangle=\frac{2}{3}\frac{k_{rad}}{m\Omega_{ce}}(F_\parallel^2+\gamma^2 F_\perp^2),\label{dipole}
\end{equation}
where $F_\parallel$ and $F_\perp$ are the parallel and perpendicular components of the force with
respect to the particle's velocity. 
The bracket $\langle\rangle$ indicates that we take the average 
of all the particles locating within $30\lambda_D$ from the pulse front, and the number of 
particles within this layer decreases with time because of the decoupling from the EM pulse. 
Since Eq. (\ref{dipole}) is proportional to $k_{rad}$, 
we plot Run A and D with $k=10^{-3}$ and $10^{-4}$ to compare with the other four RD cases.
  
For the RD cases, the radiation power is calculated using the formula
\begin{equation}
\langle P\rangle=|\ve{f}_{rad}\cdot\ve{v}|.\label{landau}
\end{equation}
We compared the result of these two formulae for the RD cases,and they matched within the linewidth.Therefore, We only show the result of Eq. (\ref{landau}) here. 

Figure \ref{fig:pow} shows the instantaneous radiation power $\langle P\rangle$ for all runs.
The initial peak around $t\Omega_{ce}=100$ is due to thermal-cyclotron emission~\cite{rybi79},
and the estimated radiation power for the NRD case quantitatively matches with the RD case.
At a later time ($t \Omega_{ce}>1000$), however, more power is irradiated in the NRD cases than 
in the RD cases, because energetic particles are continuously accelerated without 
losing their energy self-consistently in the NRD cases. In the RD cases, however, all the
energetic particles are decelerated by the radiation damping, and the instant radiation
power decreases caused by slower velocity and smaller damping force. 

\begin{figure}
\includegraphics[width=\linewidth]{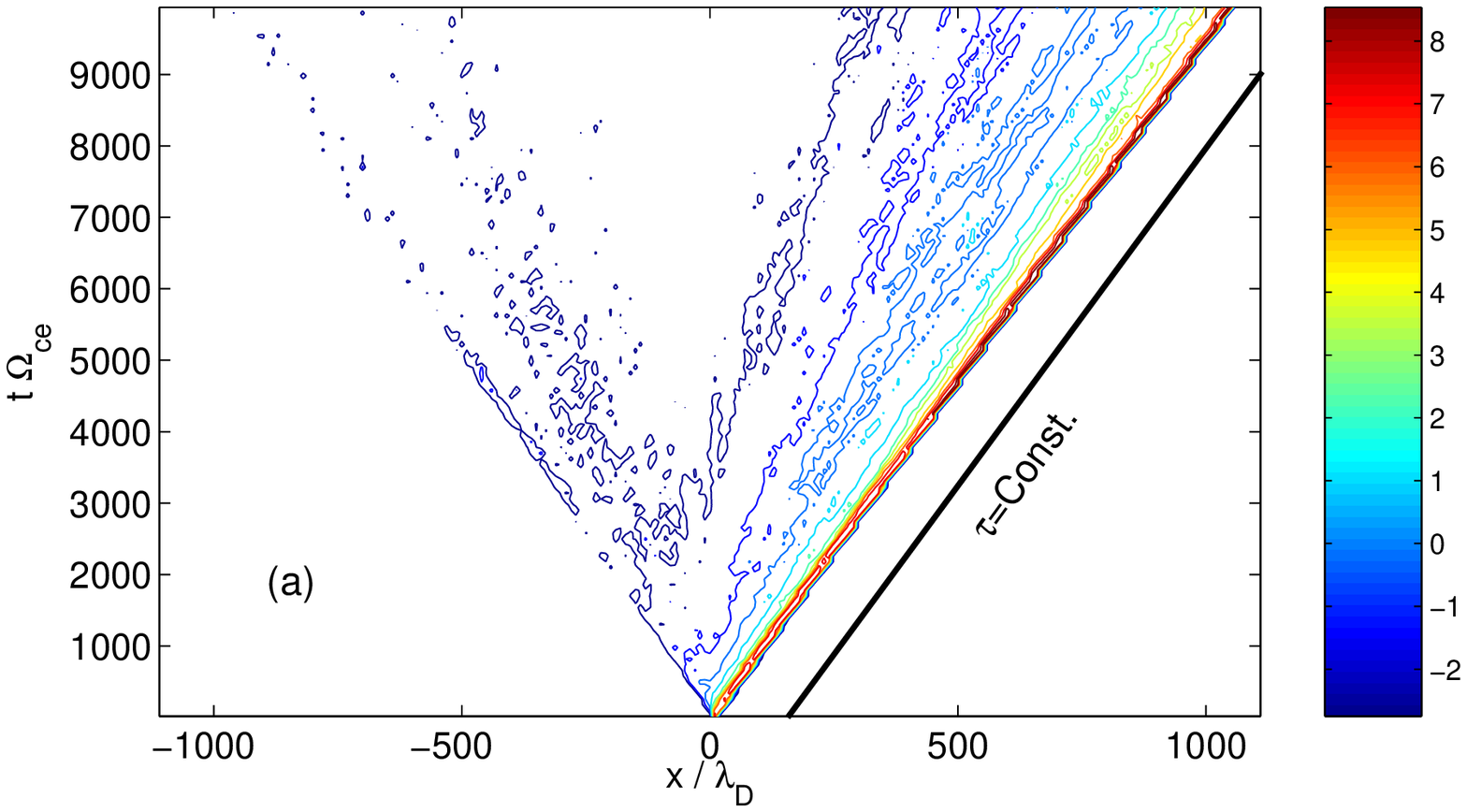}
\includegraphics[width=\linewidth]{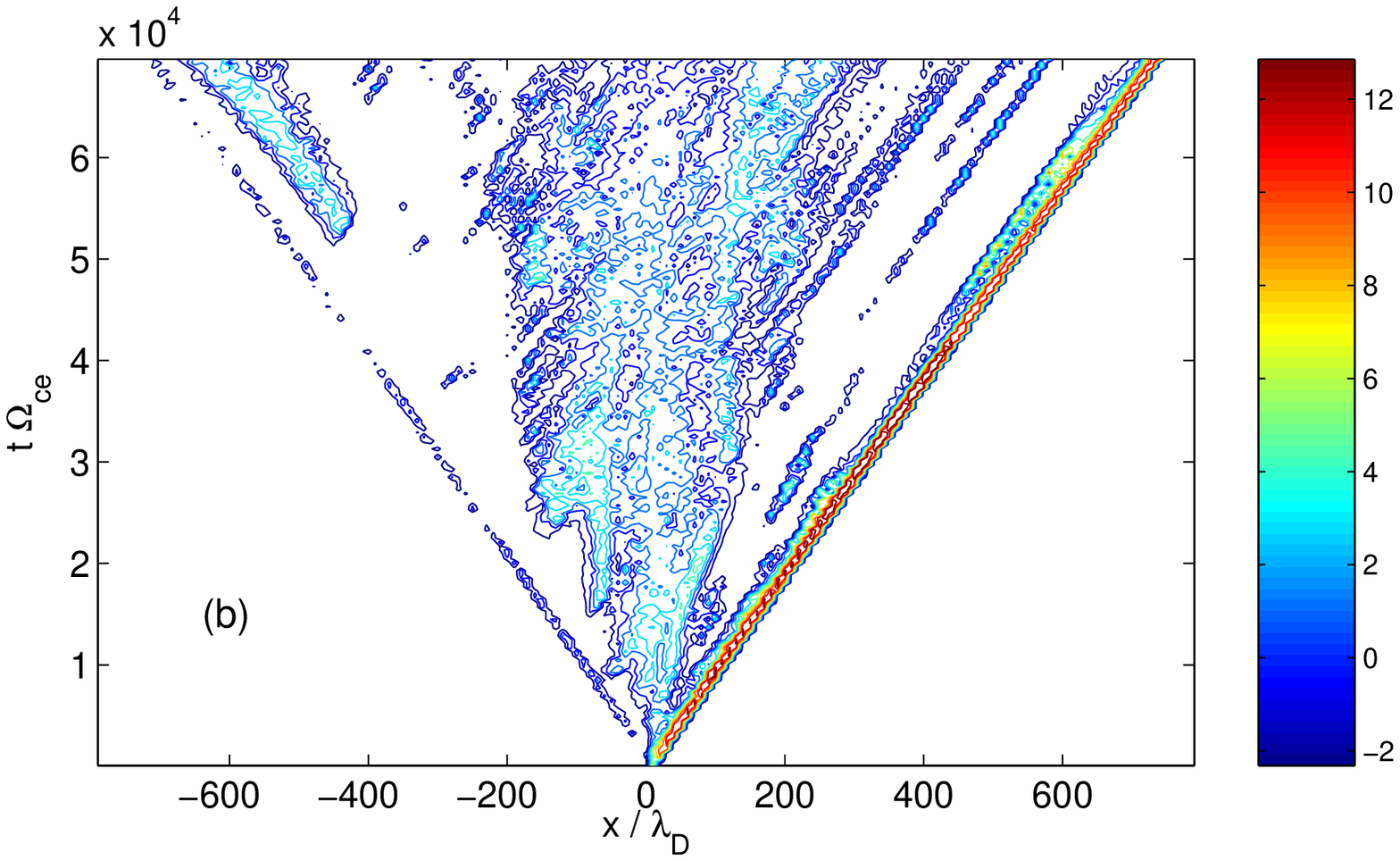}
\includegraphics[width=\linewidth]{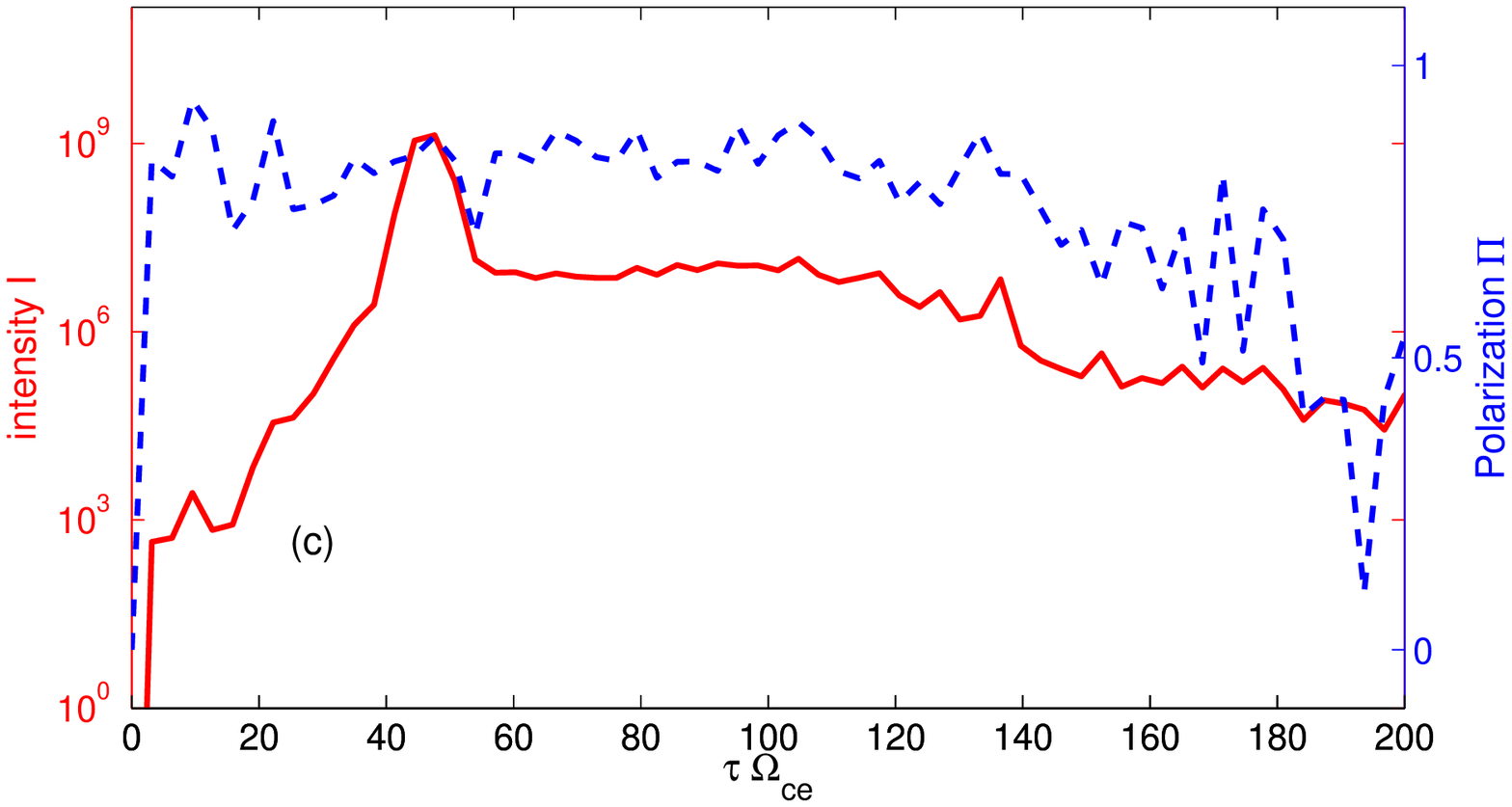}
\includegraphics[width=\linewidth]{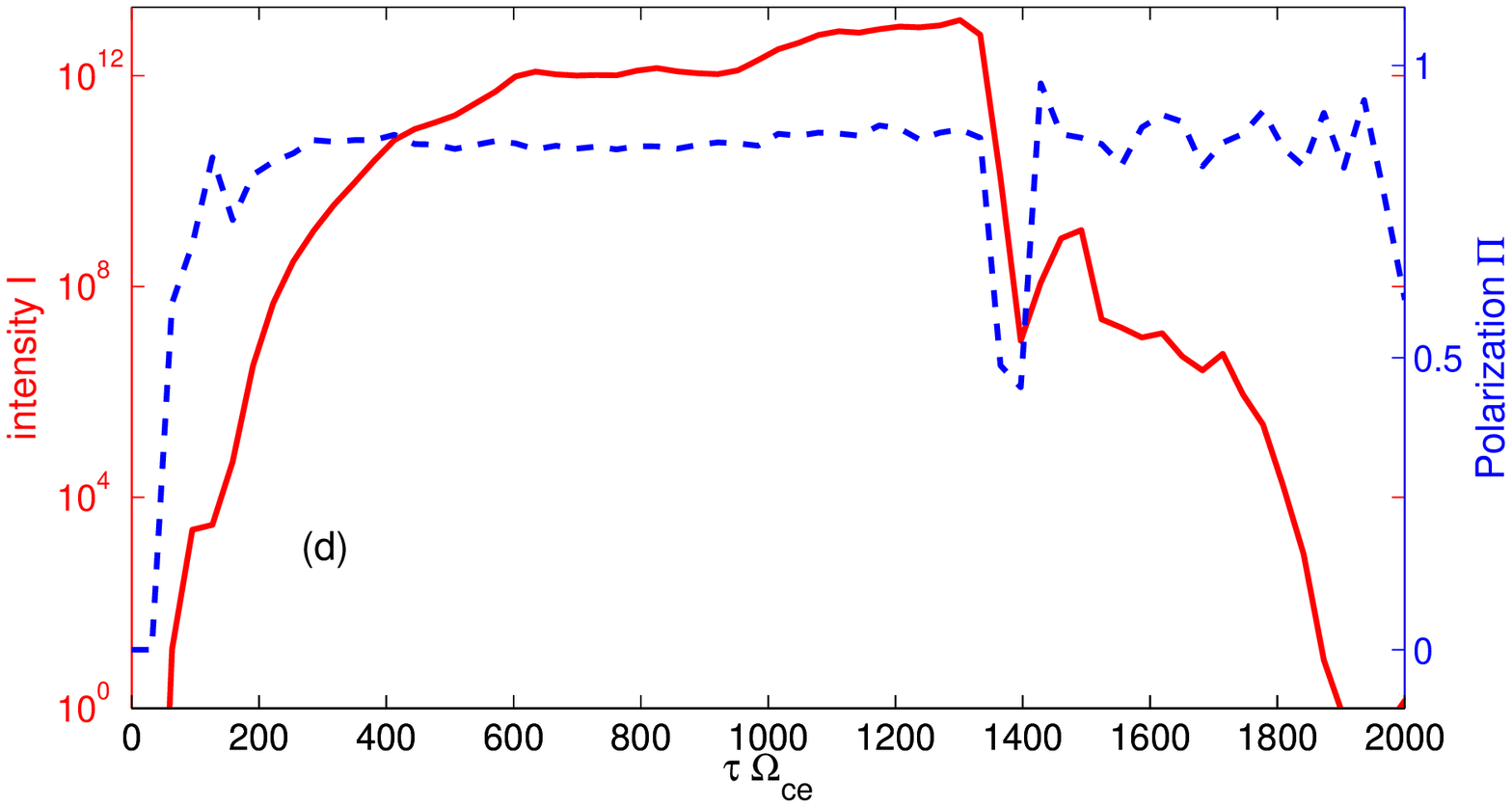}
\caption{\label{fig:int} Contour plots of instantaneous intensity $\log_{10} I$ as 
a function of local time $t$ for $\theta=\phi=0$ for (a) Run C and (b) Run F, 
and intensity (red lines, right scales) and polarization 
(blue lines, left scales) as functions of observational time $\tau$ for (c) Run C and (d) Run F. 
Intensity is in arbitrary scale. 
To obtain time dependence of intensity, instantaneous intensity from each particle
is summed up along the light cone $\tau=t-R/c=$const., shown as a black dotted line in panel (a). The
light cone moves horizontally leftward with $\tau$.}
\end{figure}

Finally, we calculate the radiation field and 
its angular dependence self-consistently using the velocity and acceleration of each particle.
Intensity $I$ and polarization $\Pi$ of the radiation received by the observer located at $\ve{x}$ 
are given by~\cite{jack75,rybi79}
\begin{equation}
I(\hat{\ve{n}},\tau)
=\sum_i\left[
    \left|\ve{E}_i\right|^2
\right]_{\mbox{ret}},\label{inten}
\end{equation}
and
\begin{eqnarray}
\!\!E_y^2(\hat{\ve{n}},\tau)\!&=&\!\sum_i\left[\left|
    \ve{E}_i\cdot\hat{\ve{y}}
\right|^2\right]_{\mbox{ret}},\nonumber\\
\!\!E_z^2(\hat{\ve{n}},\tau)\!&=&\!\sum_i\left[\left|
    \ve{E}_i\cdot\hat{\ve{z}}
\right|^2\right]_{\mbox{ret}},\nonumber\\
\!\!U(\hat{\ve{n}},\tau)\!&=&\!2\sum_i\left[
    (\ve{E}_i\cdot\hat{\ve{y}})(\ve{E}_i\cdot\hat{\ve{z}})\right]_{\mbox{ret}},\label{EyEz}\\
\!\!\Pi(\hat{\ve{n}},\tau)\!&=&\!\frac{\sqrt{(E_z^2)^2+(E_y^2)^2-2E_y^2E_z^2+U^2}}
    {E_z^2+E_y^2},
\end{eqnarray}
where 
\begin{equation}
\ve{E}_i=\frac{e}{c}\frac{\hat{\ve{n}}\times[(\hat{\ve{n}}
    -\ve{\beta}_i)\times\dot{\ve{\beta}}_i]}
    {(1-\hat{\ve{n}}\cdot\ve{\beta})^3 R},
\end{equation}
is the radiated electric field from particle $i$ located at $\ve{r}$, 
$\hat{\ve{n}}$ is a unit vector in the direction of $\ve{x}-\ve{r}(\tau)$, 
$\ve{\beta}=\ve{v}(\tau)/c$, and $\dot{\ve{\beta}}=d\ve{\beta}/dt$.
We assume that $|\ve{x}|\gg|\ve{r}|$ so that $\hat{\ve{n}}$ is parallel to $\ve{x}$.
The square brackets with a subscript "ret" mean that the quantity in the brackets is 
evaluated at the retarded time $\tau=t-R/c$, where $R=|\ve{x}-\ve{r}|$. 
To specify the direction of the observer with respect to the $x$ axis,
we introduce $\theta$ and $\phi$ as
\begin{equation}
\hat{\ve{n}}=(\cos \theta \cos \phi, \cos \theta \sin \phi, \sin \theta).
\end{equation}

In Figs.~\ref{fig:int}(a) and (b), 
the contour plot of instantaneous intensity before taking the summation 
over the retarded time is plotted, as a function of local time $t$ with 
$\phi=\theta=0$, illustrating
the ray-tracing technique used in Eqs. (\ref{inten}) and (\ref{EyEz}).
We take a sum of intensity along the light cone $\tau=t-R/c=$const., which is indicated as a black line,
up to $t\Omega_{ce}=10000$ for Run C and $t\Omega_{ce}=70000$ for Run D.
The light cone moves toward the negative $x$ direction with $\tau$, and we take $\tau=0$ when the 
pulse front reaches to the observer.
Since all the particles cross the light cone $\tau=$const. only once, we take a sum of $I$ and $\Pi$ as follows.
First, we follow the 
trajectory of each particle, and store their instantaneous intensity. 
When each particle crosses the light cone $\tau\Omega_{ce}=1$, 
we take the average of the instantaneous intensity of the particle 
over the trajectory between $\tau\Omega_{ce}=0$
and $1$, and add it to the intensity $I$. Then we continue to follow the
particle trajectory and sum up the intensity again until it hits the next
light cone. We continue the summation 
until the signal in the initial plasma thickness reaches to the observer, 
$\tau=12\lambda_D/c$, 
We also note that the intensity is extremely asymmetric because
only energetic particles accelerated in positive $x$ direction can radiate strong emission to the observer.

The time dependence of intensity and polarization with $\phi=\theta=0$ is shown in Figs.~\ref{fig:int}(c) and (d). 
If all the particle moves with the speed of light and are continuously accelerated 
toward the direction of $\hat{\ve{n}}$,
all the radiation emitted from particles should reach the observer at the same time and
the signal becomes a $\delta$-function pulse. However,
energetic particles
are bouncing back and forth within the ponderomotive potential well, and slower particles are 
dropped off to the next well, which broaden the spatial distribution and the resulting radiation duration.

Intensity $I$ is shown as solid lines in Fig.~\ref{fig:int}(c) and (d), indicating the duration time
is $20 \tau\Omega_{ce}$ for Run C  and $200\sim400\tau\Omega_{ce}$ for Run D with single peak.
Since the initially applied electromagnetic field is linearly polarized, we expect that the radiation is also strongly 
polarized, with the small depolarization coming from the initial random velocity distribution in the $y$ direction.
Polarization $\Pi$ is shown as dotted lines in Fig.~\ref{fig:int}(c) and (d),
illustrating that the radiation is strongly linear-polarized as anticipated.

\begin{figure}
\includegraphics[width=\linewidth]{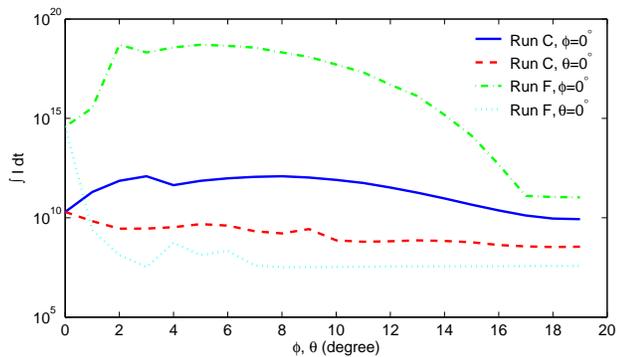}
\caption{\label{fig:itg} 
The total radiation intensity $\int I(\phi, \theta) d\tau$ for Run C and Run F as a function of the 
angle ($\phi, \theta$). }
\end{figure}

In Fig.~\ref{fig:itg}, we show the total radiation intensity $\int I(\phi, \theta) d\tau$. 
The radiation is a very short 
pulse in Fig.~\ref{fig:int} because we consider single simulation box centered at the origin only.
However, because of the periodic boundary condition in the $z$ direction, 
we should consider the multiple simulation boxes along the $z$ axis, and consider time delay
and angle difference toward the observer from each simulation box, which is very complex
even in the Cartesian coordinate.

Instead of including these geometrical effects into the intensity calculation,
we simply compare the total intensity integrated over $t$ as a function of the angle. 
Intensity peaks around $\theta=3\sim 8^\circ$ in $\phi=0$ cases (solid and dash-dot lines), corresponding to 
the direction of high energy particles in Fig.~\ref{fig:phs}. 
Intensity rapidly decreases with both $\phi$
and $\theta$, indicating radiation is strongly collimated in the $x$ direction.

Intensity distribution in the $z$ direction is due to the initial electric field acceleration,
as we discussed. Since there is no acceleration in the $z$ direction after the induced current
is formed, eventually all the momentum in the $z$ direction will be emitted away, which
narrows the intensity distribution toward positive $x$ direction. At
the same time, however, decoupling of particles from EM pulse in the $x$ direction makes
the DRPA less efficient in the $x$ direction, which widen the distribution. 
Thus, the intensity in the $x-z$ plane is always distributed over a finite angle,
but the distribution in much later time is still an open question.  

\section{SUMMARY\label{sec:sec5}}
In summary, we observed the self-consistent radiation damping effect on the interaction of the DRPA
with electron-positron plasma via a relativistic PIC simulation. We have found that field and particle
energies are transferred to radiation, and the coupling between the field and particles becomes less efficient 
with larger radiation damping. Comparison with the non-radiative case showed that radiation
damping force decelerates the energetic particles accelerated by the DRPA, and resulting radiation
power is smaller in the radiative case. The radiation field is strongly linearly polarized both in
weak and strong magnetic field cases, which may be detectable by $\gamma$-ray burst observations or
laser experiments as an indication of the DRPA mechanism.
The simulations shown here are too short to see multiple peaks in an intensity 
time profile as seen in GRB observations, and to determine the final radiation 
pattern after particles are completely decoupled from EM pulse. 
These two questions remain as future problems.

\bigskip 
\begin{acknowledgments}
This research is partially supported by NASA Grant No. NAG5-9223 
and LLNL contract nos. B528326 and B541027. 
The authors wish to thank ILSA, LANL, B. Remington and S. Wilks 
for useful discussions.
\end{acknowledgments}

\bigskip 
\bibliography{apsprl}





\end{document}